\documentclass[slac_one]{revtex4}
\usepackage{graphicx}
\usepackage{fancyhdr}
\usepackage{amsmath,amssymb}

\begin{document}

\title{Light Charged and CP-odd Higgses in MSSM-like Models} 

%

\author{Radovan Derm\' \i\v sek}
\affiliation{Physics Department, Indiana University, Bloomington, IN 47405}
%

\begin{abstract}
 We study the Higgs sector of supersymmetric models containing two Higgs doublets with a light MSSM-like CP odd Higgs, $m_A \lesssim 10$ GeV, and $\tan \beta \lesssim 2.5$. 
In this scenario
all Higgses resulting from two Higgs doublets: light and heavy CP even Higgses, $h$ and $H$, the CP odd Higgs, $A$, and the charged Higgs, $H^\pm$, could have been produced at LEP or the Tevatron, but would have escaped detection because they decay in modes that have not been searched for or the experiments are not sensitive to. Especially $H \to ZA$ and $H^\pm \to W^{\pm \star} A$ with $A \to c \bar c, \tau^+ \tau^-$
present an opportunity to discover some of the Higgses at LEP, the Tevatron and also at B factories. 
In addition, the $2.8 \sigma$ excess of the branching ratio $W \to \tau \nu$ with respect to the other leptons measured at LEP correlates well with the existence of the charged Higgs with properties typical for this scenario.
Dominant  $\tau$- and $c$-rich 
decay products of all Higgses require modified strategies for their discovery at~the~LHC.
\end{abstract}

\maketitle

\thispagestyle{fancy}

In the MSSM it is typically the case that the mass  of one of the CP even Higgses scales with the mass of the CP odd Higgs and it is the other CP even Higgs (the mass of which is related to the mass of the Z boson) that is standard model like in its coupling to the Z boson. This generic feature is not significantly modified by  including radiative corrections from superpartners. An exception to this rule is the region with $m_A \ll m_Z$ and $\tan \beta \lesssim 2.5$ which is the focus of this talk~\cite{Dermisek:2008id}.

\begin{figure}[t]
\includegraphics[width=1.5in]{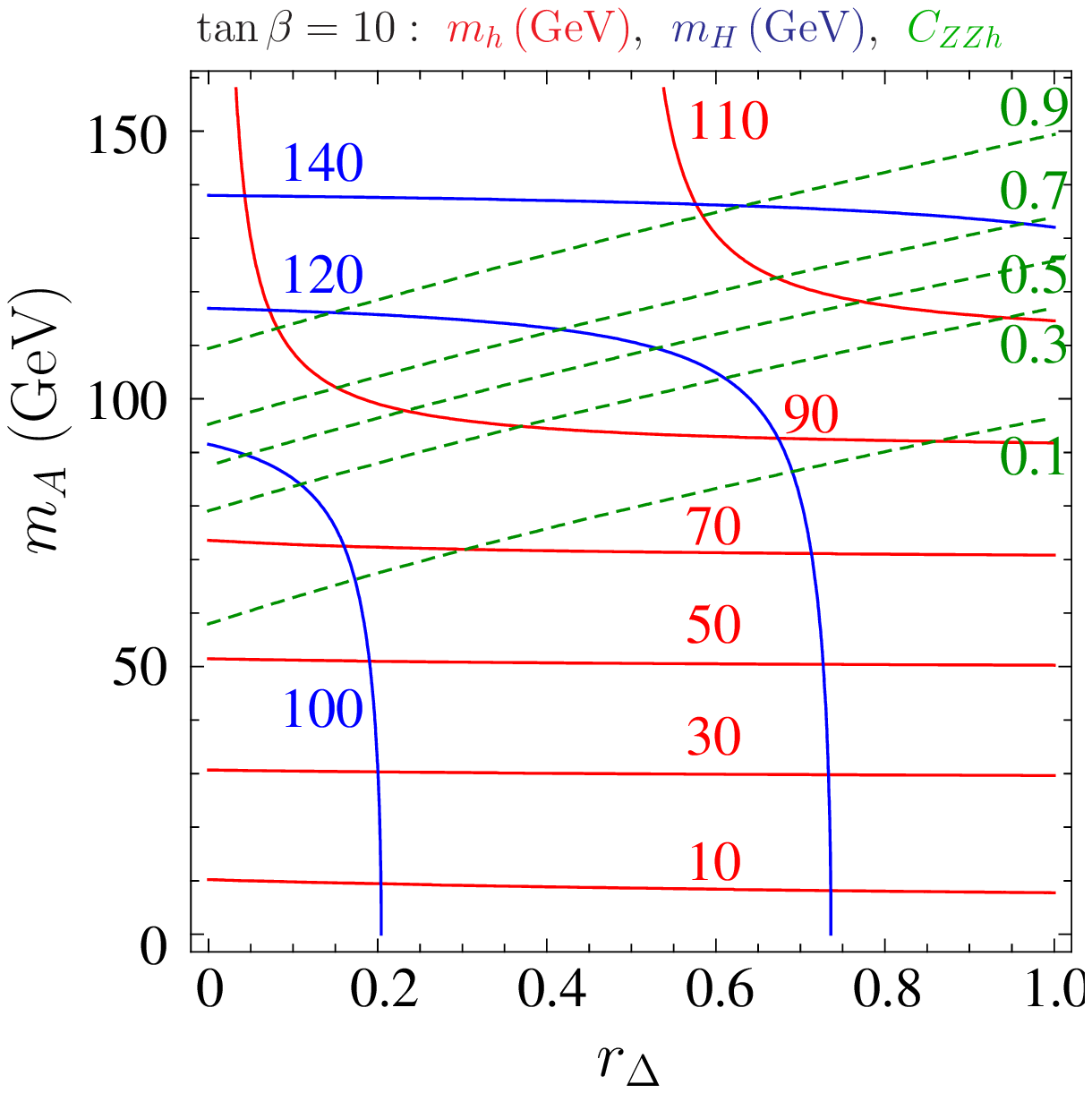}
\includegraphics[width=1.5in]{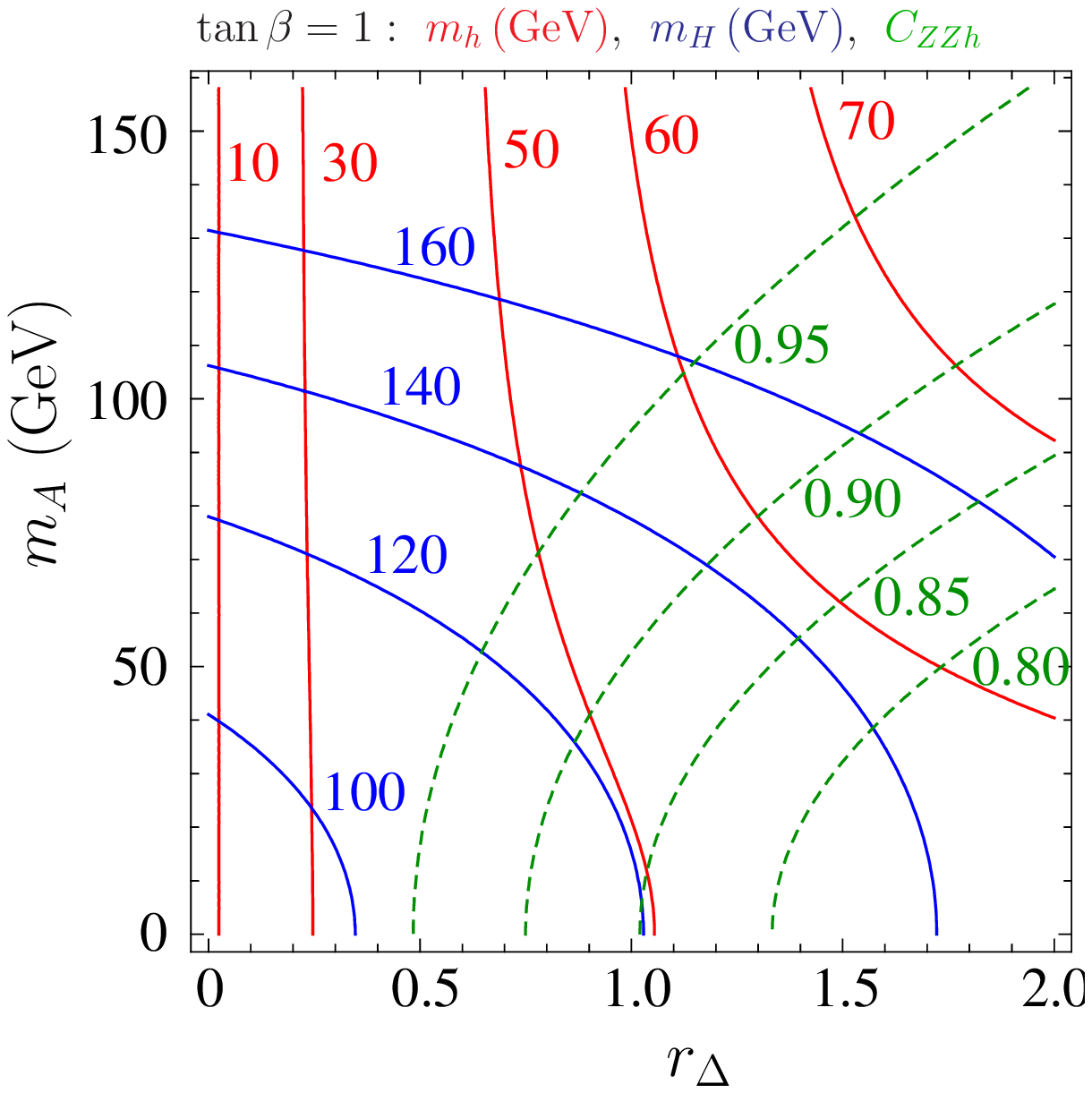}
\caption{The mass of the light $h$ (red), heavy $H$ (blue) CP even Higgs bosons and $C_{ZZh}$ (green) as a function of $m_A$ and the radiative correction from superpartners, $r_\Delta = \Delta/m_Z^2$, for $\tan \beta = 10$ (left) and $\tan \beta = 1$ (right).} 
\label{fig:higgs}
\end{figure}

For $m_A \ll m_Z$ as $\tan \beta $ approaches 1 the situation  described above dramatically changes, see Fig.~\ref{fig:higgs}. 
The light CP even Higgs boson becomes SM-like, $C_{ZZh} \equiv g^2_{ZZh}/g^2_{ZZh_{SM}} 
\simeq 1$, and although
it is massless at the tree level, it will receive a contribution from superpartners and the tree level relation between the light CP even and CP odd Higgses, $m_h < m_A$ is typically not valid. Even for modest superpartner masses the light CP even Higgs boson will be heavier than $2m_A$ and thus $h \to AA$ decay mode is open and generically dominant. For small $\tan \beta$ the width of $A$ is shared between $\tau^+ \tau^-$ and $c \bar c$ for $m_A < 2m_b$ and thus the width of $h$ is spread over several different final states, $4\tau$, $4c$, $2\tau 2c$ and highly suppressed $b \bar b$ and thus the LEP limits in each channel separately are highly weakened \cite{4tau}.

Although this scenario is ruled out in the MSSM 
we will argue that it 
easily viable in simple extensions of the MSSM~\cite{Dermisek:2008id}.
Any model that can increase the mass of the SM-like Higgs above the decay-mode-independent limit, $ 82$ GeV, will generically satisfy all other experimental limits.
We will see that decay modes of the heavy CP even Higgs (that also turns out to be within the reach of LEP) are modified in this region and even the charged Higgs boson can be below LEP or Tevatron limits due to decay modes that have not been searched for.

For $\tan \beta = 1.01$, $\mu = 100$ GeV, $m_A = 8$ GeV and varying soft susy breaking scalar and gaugino masses between 300 GeV and 1 TeV and mixing in the stop sector, $X_t/m_{\tilde t}$, between 0 and -2, we typically find: $m_h \simeq 38 - 56$ GeV with $g_{ZZh}/g_{ZZh_{SM}} \simeq 0.84 - 0.97$, $m_H \simeq 108 - 150$ GeV and  $m_{H^\pm} \simeq 78 - 80$ GeV (the charged Higgs mass is generically close to $m_W$ in this scenario as a consequence of $m_{H^\pm}^2 = m_W^2 + m_A^2 \simeq m_W^2$). The dominant branching ratios of the light CP even Higgs are typically:
\vspace{-0.1cm}
\begin{equation}
B (h \to A A, \; b \bar b)  \; \simeq \;  90 \%,  \; 10 \% 
\label{eq:Bh}
\end{equation}
\vspace{-0.2cm}
with 
\vspace{-0.1cm}
\begin{equation}
B (A \to \tau^+ \tau^-, \; c \bar c, \; gg )  \; \simeq \;  50 \%, \; 40 \%, \; 10\% ,
\label{eq:BA}
\end{equation}
for $2m_\tau \lesssim m_A \lesssim 10$ GeV. Branching ratios of the Heavy CP even Higgs vary with SUSY spectrum. 
For 1 TeV SUSY and $X_t/m_{\tilde t} = 0$ we find:
\vspace{-0.1cm}
\begin{equation}
B (H \to ZA, \; A A, \; hh, \; b \bar b)  \; \simeq \;  37 \%, \; 34 \%, \; 28 \% , \; 0.4 \%
\label{eq:BH}
\end{equation}
Finally, the dominant branching ratios of the charged Higgs are:
\vspace{-0.1cm}
\begin{equation}
B (H^+ \to W^{+ \star } A, \; \tau^+ \nu, \; c \bar s)  \; \simeq \;  70 \%, \; 20 \%, \; 10 \% . 
\label{eq:BHpm}
\end{equation}
For discussion of experimental constraints let us also include branching ratios of the top quark:
\vspace{-0.1cm}
\begin{equation}
B(t \to H^+ b) \simeq 40 \%, \quad B(t \to W^+ b) \simeq 60 \% .
\label{eq:Bt}
\end{equation}
These results (except (\ref{eq:BH})) are not 
very sensitive to superpartner masses nor 
the mass of the CP odd Higgs as far as $m_A < 2m_b$. Increasing $\tan \beta$ to $2.5$ 
only the following branching ratios significantly change: $B (A \to \tau^+ \tau^-, gg )  \simeq  90 \%,  10\%$, $B (H^+ \to W^{+ \star } A, \tau^+ \nu)  \simeq  35 \%, 65 \%$ and $B(t \to H^+ b, W^+ b) \simeq 10 \%, 90 \%$.

The experimental constraints on this scenario are discussed in detal in Ref.~\cite{Dermisek:2008id}.
The mass of the light CP even Higgs is the only problematic prediction.
The decay-mode independent search from OPAL sets the limit on the Higgs mass by looking only for reconstructed Z boson decaying leptonically and excludes $m_h < 82$ GeV for $C_{ZZh} = 1$.
There are however various ways to increase the mass of the SM-like Higgs boson in extensions of the MSSM. A simple
possibility is to consider singlet extensions of the MSSM containing $\lambda S H_u H_d$ term in the superpotential.
It is known that this term itself contributes $\lambda^2 v^2 sin^2 2 \beta$, where v = 174 GeV, to the mass squared of
the CP even Higgs
 and thus can easily push the Higgs mass above the decay-mode independent limit, 
$82$ GeV.
 Note, this contribution
is maximized for $\tan \beta \simeq 1$. 
In this talk we assume that a possible extension does not significantly alter the two Higgs
doublet part of the Higgs sector besides increasing the Higgs mass above the decay-mode independent limit.\footnote{ 
For example, the next-to-minimal supersymmetric model (NMSSM) has a limit in which it resembles the MSSM . Indeed, in the NMSSM the scenario with a light MSSM-like CP odd Higgs and small $\tan \beta$ is viable and has all the features of the MSSM in this limit~\cite{NMSSM_small_tb}. It should be stressed however that this scenario is not limited to singlet extensions of the MSSM and it would be viable in many models beyond the MSSM that increase the mass of the SM-like Higgs boson.}

The rest of the Higgs spectrum is basically not constrained at all in this scenario. 
The heavy CP even and the CP odd Higgses could have been produced at LEP in $e^+ e^- \to H A$ but they would avoid detection because $H$ dominantly decays to $ZA$ - the mode  that has not been searched for. The searches in various final states of $AH \to AAA$ are either not sensitive to or were not done in the range of masses typical in our scenario.

The light CP odd Higgs might be also within the reach of current B factories where it can be produced  
in Upsilon decays, $\Upsilon \to A \gamma$. 
This was recently suggested in the framework of the
next-to-minimal supersymmetric model (NMSSM) with a light CP odd Higgs boson being mostly the singlet~\cite{Dermisek:2006py} (in our scenario $A$ is doublet-like) and it overlaps with searches for lepton non-universality in $\Upsilon$ decays~\cite{Sanchis-Lozano}. It is advantageous to look for a light CP odd Higgs in $\Upsilon (1S, 2S, 3S)$ since these states cannot decay to B mesons and thus the $A \gamma$ branching ratio is enhanced.
Predictions for the branching ratio $B(\Upsilon \to A \gamma)$ for $\tan \beta = 1$ can be readily (although only approximately) obtained from the results of Ref.~\cite{Dermisek:2006py} taking $\tan \beta \cos \theta_A \simeq 1$ ($\cos \theta_A$ is the doublet component  which is 1 for MSSM-like CP odd Higgs).
Recent 
CLEO  limits~\cite{CLEOprelim} do not constrain our scenario for $m_A \gtrsim 7.5$ GeV.

The strongest limits on the charged Higgs from both LEP and the Tevatron assume $H^+ \to \tau^+ \nu$ or $H^+ \to c \bar s$. In our scenario, the dominant decay mode of the charged Higgs is
$H^\pm \to W^{\pm \star} A$ with $A \to c \bar c$ or $\tau^+ \tau^-$. This decay mode was never searched for. The current limits do not rule out  the charged Higgs as light as 75 GeV assuming decay modes typical for our scenario and $B(t \to H^+ b) \lesssim 40\%$ is currently not excluded. 

As we discuss next, the charged Higgs with properties typical for this scenario could explain the deviation from lepton universality in $W$ decays measured at LEP~\cite{Dermisek:2008dq}.


From the combined results of LEP collaborations on the 
leptonic branching ratios of the W boson an excess of the branching ratio 
$W \to \tau \nu$ with respect to the other leptons is evident~\cite{:2004qh}.
While branching ratios of $W \to e \nu$ and $W \to \mu \nu$ perfectly agree with 
lepton universality,
\vspace{-0.1cm}
\begin{equation}
B(W \to \mu \nu) / B(W \to e \nu) = 0.994 \pm 0.020,
\end{equation}
the branching fractions in $\tau$ with respect to $e$ and $\mu$ differ by more than $2\sigma$.
The ratio between the tau fraction and the average of electron and muon fractions,
\vspace{-0.1cm}
\begin{eqnarray}
&R_{\tau/l} \equiv 2B(W \to \tau \nu) / (B(W \to e \nu)+B(W \to \mu \nu)) ,& \nonumber \\
&R_{\tau/l}^{exp} = 1.073 \pm 0.026,&
\label{eq:Rtau_l_exp}
\end{eqnarray}
results in a poor agreement, at the level of 2.8 standard deviation, with lepton universality.

The $WW$ pair production cross section, $\sigma_{W^+W^-}$, at LEP is about 17 pb at the center of mass energy $\sqrt{s} = 200$ GeV and $W^\pm$ decay equally (in the SM) to each generation of leptons with branching ratio of $10.6 \%$. 
Since charged Higgs pair production cross section, $\sigma_{H^+H^-}$, is about 160 fb for $m_{H^\pm} \simeq m_{W^\pm}$, about two orders of magnitude smaller than $\sigma_{W^+W^-}$, and charged Higgs may decay to $\tau \nu$ with significantly larger branching fraction than $W$ (depending on the parameter space) already a naive estimate suggests that a charged Higgs with mass close to the mass of the $W$ boson can easily contribute to the measurement of lepton universality at LEP at the level indicated by the experimental result~(\ref{eq:Rtau_l_exp}).

Lepton universality in $W$ decays was measured also at 
the Tevatron.
CDF~\cite{Safonov:2004zv} is looking at inclusive W production
and the ratio
 $Br(W \to \tau \nu)/Br(W \to e \nu)  = 0.99 \pm 0.04(stat) \pm 0.07(syst)$
agrees with lepton universality.
W bosons are produced in $p \bar p$ interactions dominantly through the Drell-Yan process
The production cross section of a single charged Higgs from first-generation quarks is obviously negligible
 and thus the charged Higgs is not expected to affect lepton universality 
in this measurement.

Direct production of the charged Higgs boson with mass close to the mass of W boson is a unique way to explain the deviation from lepton universality in W decays at LEP and agreement with lepton universality in W decays measured at the Tevatron.\footnote{The possibility of a charged Higgs explanation of the lepton non-universality in W boson decays was also    discussed in a different framework by J. H. Park~\cite{Park:2006gk}.} Any possible alternative explanation by new physics that would modify the $W\tau \nu$ vertex through loop corrections would necessarily predict the deviation from lepton universality at both LEP and the Tevatron. 


\begin{figure}[t]
\hspace{-1.cm}
\includegraphics[width=2.in]{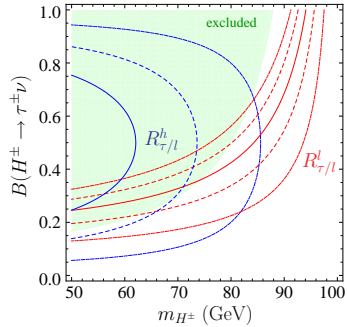}
\caption{$R_{\tau/l}^l$ (red) and $R_{\tau/l}^h$ (blue) as a function of $m_{H^\pm}$ and $B(H^\pm \to \tau \nu)$ for $\sqrt{s} = 200$ GeV. Solid lines represent $R_{\tau/l}^l, R_{\tau/l}^h = R_{\tau/l}^{exp} = 1.073$ and dashed and dotted lines indicate $1\sigma = \pm 0.026$ and $2\sigma$ ranges. Shaded region is excluded by LEP searches for the charged Higgs boson, assuming $B(H^\pm \to \tau \nu) = 1$. Other limits apply for $m_{H^\pm} \lesssim 75$ GeV that are not easy to implement in the plot (see the text).}
\label{fig:RWtaunu}
\end{figure}

Charged Higgs can contribute in the fully leptonic
$\tau \nu \tau \nu$ and semi-leptonic $\tau \nu + hadrons$ 
channels.
Its contribution in the $\tau \nu \tau \nu$ channel would manifest itself in the excess of 
$\tau \nu \tau \nu$ events compared to $l \nu l \nu$, $l = e,\mu$ events and 
would be attributed to the
larger branching ratio of $W \to \tau \nu$ compared to $W \to l \nu$,  $l = e,\mu$.
This increase is given by\footnote{Charged Higgs can contribute directly only to $\tau \nu \tau \nu$ channel and not to mixed 
$\tau \nu l \nu$,  $l = e,\mu$ channels. However if $\tau$ decays leptonically the 
efficiency of an  $W \to \tau \nu$ event to pass as a $W \to l \nu$ event
 is not small and so the charged Higgs production would effectively contribute 
to both $\tau \nu \tau \nu$ and mixed $\tau \nu l \nu$ channels. 
For this reason the prediction of $R_{\tau/l}^l$ should be treated only as an estimate of the
effect of the charged Higgs on lepton non-universality in $W$ decays.}
\begin{equation}
R_{\tau/l}^l = \sqrt{1 + \frac{\sigma_{H^+ H^-} B(H^+ \to \tau^+ \nu)^2}
{\sigma_{W^+ W^-} B(W^+ \to l^+ \nu)^2}}.
\end{equation}

In a similar way the contribution to the $\tau \nu + {\rm hadrons}$ final state that 
would be attributed to the
larger branching ratio of $W \to \tau \nu$ compared to $W \to l \nu$,  $l = e,\mu$
can be roughly estimated by
\begin{equation}
R_{\tau/l}^h = 1 + \frac{\sigma_{H^+ H^-} B(H^+ \to \tau^+ \nu) B(H^+ \to hadrons)}
{\sigma_{W^+ W^-} B(W^+ \to l^+ \nu)B(W^+ \to hadrons)}
\label{eq:Rtaulhad}
\end{equation}
where we take
$B(H^+ \to hadrons) \simeq 1- B(H^+ \to \tau^+ \nu)$.
Due to complicated final states of $H^\pm$ this should be considered only as a rough estimate~\cite{Dermisek:2008dq}.

In Fig.~\ref{fig:RWtaunu} we show $R_{\tau/l}^l$ (red) and $R_{\tau/l}^h$ (blue) as a function of $m_{H^\pm}$ and $B(H^\pm \to \tau \nu)$ for $\sqrt{s} = 200$ GeV. Solid lines represent $R_{\tau/l}^l, R_{\tau/l}^h = R_{\tau/l}^{exp} = 1.073$ and dashed and dotted lines indicate $1\sigma = \pm 0.026$ and $2\sigma$ ranges. Shaded region is excluded by LEP searches for the charged Higgs boson, assuming $B(H^\pm \to \tau \nu) = 1$. Other limits apply for $m_{H^\pm} \lesssim 75$ GeV  as we discussed before 
but these are not easy to implement in the plot because they depend on other parameters, e.g. $\tan \beta$. We see that the charged Higgs with mass $75 - 85$ GeV and $B(H^+ \to \tau^+ \nu) \simeq 20 - 60 \%$ has the right properties to explain the measured deviation from lepton universality in $W$ decays. The properties of the charged Higgs favored by the $R_{\tau/l}^{exp} $ are exactly those found in the  $m_A \ll m_W$, $\tan \beta \lesssim 2.5$ scenario~(\ref{eq:BHpm}).

Clearly the search for the charged Higgs including the dominant  $W^\star A$ with $A \to c \bar c$ or $\tau^+ \tau^-$ decay modes at LEP and especially at the Tevatron with currently available much larger data sample is very desirable.

%



%




\begin{thebibliography}{99}   


\bibitem{Dermisek:2008id}
  R.~Dermisek,
  arXiv:0806.0847 [hep-ph].
  

\bibitem{4tau}
This feature is similar to $h \to 2a \to 4\tau$ scenario discussed in the context of NMSSM:
  R.~Dermisek and J.~F.~Gunion,
  Phys.\ Rev.\ Lett.\  {\bf 95}, 041801 (2005);
%
  Phys.\ Rev.\  D {\bf 73}, 111701 (2006);
  %
  Phys.\ Rev.\  D {\bf 75}, 075019 (2007);
%
  Phys.\ Rev.\  D {\bf 76}, 095006 (2007).
%
%
  For a review 
  see also, S.~Chang {\it et al.}, 
  arXiv:0801.4554 [hep-ph].

  

\bibitem{NMSSM_small_tb}
R.~Dermisek and J.~F.~Gunion, in preparation.


\bibitem{Dermisek:2006py}
  R.~Dermisek, J.~F.~Gunion and B.~McElrath,
  Phys.\ Rev.\  D {\bf 76}, 051105 (2007)
  [arXiv:hep-ph/0612031].

\bibitem{Sanchis-Lozano}
%
  M.~A.~Sanchis-Lozano,
  Mod.\ Phys.\ Lett.\  A {\bf 17}, 2265 (2002);
%
  Int.\ J.\ Mod.\ Phys.\  A {\bf 19}, 2183 (2004);
%
  E.~Fullana and M.~A.~Sanchis-Lozano,
  Phys.\ Lett.\  B {\bf 653}, 67 (2007).

  \bibitem{CLEOprelim}
  S.~Stone, 
FPCP 08, Taipei, Taiwan, May 5-9, 2008.

\bibitem{Dermisek:2008dq}
  R.~Dermisek,
  arXiv:0807.2135 [hep-ph].

\bibitem{:2004qh}
    [LEP Collaborations],
  arXiv:hep-ex/0412015.


\bibitem{Safonov:2004zv}
  A.~Safonov  [the CDF collaboration],
  Nucl.\ Phys.\ Proc.\ Suppl.\  {\bf 144}, 323 (2005).
  

\bibitem{Park:2006gk}
  J.~h.~Park,
  JHEP {\bf 0610}, 077 (2006).

\end{thebibliography}
\end{document}